\documentclass{article}
\usepackage{waspaa21,amsmath,graphicx,url,times}
\usepackage{color}
\usepackage{amsfonts}
\usepackage{multirow}
\usepackage{epstopdf}
\usepackage{CJKutf8}

\title{Multi-Scale Attention Neural Network for Acoustic Echo Cancellation}

\name{Lu Ma\sthanks{This work was supported by National Key R$\&$D Program of China, under Grant No. 2020AAA0104500. The corresponding author is Lu Ma. Email: malu6@tal.com, iamroad@163.com},
      Song Yang,
      Yaguang Gong,
      Zhongqin Wu}
\address{TAL Education Group, Beijing, China\\
         \{malu6,yangsong1,gongyaguang,wuzhongqin\}@tal.com}

\begin{document}

\ninept
\maketitle

\begin{sloppy}

\begin{abstract}
  Acoustic Echo Cancellation (AEC) plays a key role in speech interaction by suppressing the echo received at microphone introduced by acoustic reverberations from loudspeakers. Since the performance of linear adaptive filter (AF) would degrade severely due to nonlinear distortions, background noises, and microphone clipping in real scenarios, deep learning has been employed for AEC for its good nonlinear modelling ability. In this paper, we constructed an end-to-end multi-scale attention neural network for AEC. Temporal convolution is first used to transform waveform into spectrogram. The spectrograms of the far-end reference and the near-end mixture are concatenated, and fed to a temporal convolution network (TCN) with stacked dilated convolution layers. Attention mechanism is performed among these representations from different layers to adaptively extract relevant features by referring to the previous hidden state in the encoder long short-term memory (LSTM) unit. The representations are weighted averaged and fed to the encoder LSTM for the near-end speech estimation. Experiments show the superiority of our method in terms of the echo return loss enhancement (ERLE) for single-talk periods and the perceptual evaluation of speech quality (PESQ) score for double-talk periods in background noise and nonlinear distortion scenarios.
\end{abstract}

\begin{keywords}
acoustic echo cancellation, AEC, temporal convolution network, multi-sacle, attention
\end{keywords}

\section{Introduction}
\label{sec:intro}

Acoustic echo will arise when the microphone at the near-end picks up the loudspeaker's sound plus its reverberation and heard by the speaker itself at the far-end. This makes it very annoying in speech, audio and acoustic applications, such as teleconferencing, hands-free telephony, and mobile communication. It has been received significant attention for decades \cite{Sondhi,Benesty}. Since there is a reference signal named the far-end, adaptive filters (AF) are always employed for acoustic echo cancellation (AEC) \cite{Breining}. Several classical AF algorithms have been proposed, such as least mean square (LMS) \cite{FLMS}, normalized LMS (NLMS) \cite{NLMS}, block LMS (BLMS) \cite{BLMS}, and etc. Among them the NLMS algorithm family is most widely used due to its relatively robust performance and low complexity, such as the frequency domain block adaptive filter (FDBAF) \cite{PBFDAF} and the multidelay block frequency domain (MDF) adaptive filter \cite{MDF}.

However, since there would be non-linear components on the audio devices, nonlinear distortion would be introduced into the acoustic echo in practices.
This will lead to considerable residual echo when employing AF methods. To overcome this problem, numeric methods have been proposed, such as Nonlinear AEC (NLAEC) method where a set of nonlinear basis functions are used for AEC \cite{Stenger,Kuech,Carini} and nonlinear post-filtering method where an additional nonlinear processing module is cascaded for residual echo suppression \cite{Gustafsson,Turbin,Bendersky,Schwarz}. Recently, since its great potential in speech processing tasks, neural network (NN) has been used for AEC, such as NN-based post-filtering method where NN is used for residual echo suppression instead of conventional post-filters \cite{MaLu}, NN-based NLAEC method where NN is used for modeling nonlinear echo \cite{Halimeh1}, separation-based method where source separation with an additional information from the far-end \cite{DeLiang} is used for AEC, NN-based adaptive filtering method where the structure of AF is adopted for designing AEC network \cite{Fazel,Fazel1}.

In this paper, we constructed an end-to-end framework for AEC. 1--D (D denotes $dimension$) temporal convolution is first used to transform waveform into spectrogram, performing like a short-time Fourier transformation (STFT) with high resolution. The spectrograms of the far-end reference and the near-end mixture are concatenated, and then fed to a temporal convolution network (TCN) with stacked dilated convolution layers \cite{ref_tcn} for considering different scale features.
Attention mechanism \cite{ref_attention} is performed among these representations from different layers to adaptively extract relevant features by referring to the previous hidden state in the encoder long short-term memory (LSTM) unit. These representations are multiplied with the corresponding attention weights and added together, then fed to the encoder LSTM for the near-end speech estimation. This idea is inspired by \cite{ref_darnn} where an input attention mechanism is used to adaptively extract relevant driving series, but it is different with that of \cite{Fazel}\cite{xielie} where attention is used among consecutive time-series features extracted by gate recurrent unit (GRU) or long-short term memory (LSTM). In our scheme, since dilated convolutions are stacked, the perceptive field is increasing with more new data being covered, resulting in different scales. Attention is used to consider the data from different scales.

The remainder of this paper is organized as follows. Section 2 provides the details of our proposed network structure. Section 3 presents our experimental results. And finally, the summarization and discussions are given in Section 4.

\begin{figure*}[htb]
\centering
\includegraphics[scale=0.7]{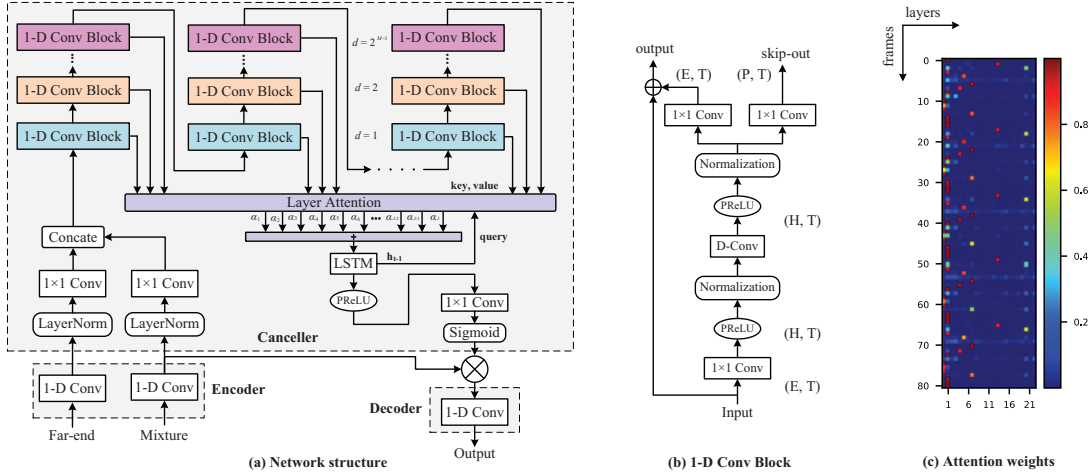}
\caption{Network architecture.}
\label{fig:structure}
\end{figure*}

\section{Network Architecture}
\label{sec:structure}
Inspired by the adaptive filter method where echo is first estimated by considering the far-end and the mixture as inputs, and then subtracted the estimated echo from the mixture. Here, the near-end speech is estimated by considering both the mixture spectrogram and the near-end spectrogram. The network structure is shown in Fig.~\ref{fig:structure}. It is constructed by an encoder, an canceller (which is constructed by a feature extractor and a mask generator) and a decoder. Encoder modules are used to transform short segments of the input waveform into their corresponding representations in an intermediate feature space. These representations are then used in the canceller to estimate masks for the mixture signal at each time step. The near-end waveform is then reconstructed by transforming the masked representation using a decoder module.

\subsection{Encoder and Decoder}
\label{sec:encoder}
The input audio is divided into overlapping segments of length $L$ samples. It is represented by $x_{k} \in \mathbb{R}^{1 \times L}$, where $k=1, \ldots, {T}$ denotes the segment index and ${T}$ denotes the total number of segments. $x_{k}$ is transformed into a $N$-dimensional representation, by a 1--D convolution operation $\mathbf{w} \in \mathbb{R}^{1 \times N}$ (denoted by $1$-$D$ $Conv$). It is formulated by a matrix multiplication as ${\mathbf{w}}=\mathcal{H}(\mathrm{\mathbf{x}} \mathbf{U})$, where $\mathbf{U} \in \mathbb{R}^{N \times L}$ contains $N$ vectors (encoder basis functions) with length $L$ for each, $\mathcal{H}(\cdot)$ is the rectified linear unit (ReLU) function~\cite{ref_relu} to ensure non-negative of the representation.

The decoder reconstructs the waveform from representation using a 1--D transposed convolution operation. It is reformulated as matrix multiplication as $\hat{\mathbf{x}}=\mathbf{w} \mathbf{V}$, where $\hat{\mathbf{x}} \in \mathbb{R}^{1 \times L}$ is the reconstruction of $\mathbf{x}$ and the rows in $\mathbf{V} \in \mathbb{R}^{N \times L}$ are the decoder basis functions, each with length $L$ samples. The overlapping reconstructed segments are summed together
to get the final waveforms.

\subsection{Multi-Scale Feature Extractor}
\label{sec:masking}
Layer normalization is firstly used to ensure that the calculation is invariant to the input scaling. The pointwise convolution (denoted by $1$$\times $$1$-$Conv(\cdot)$)~\cite{ref_depthwise} is used as bottleneck to compress the input number of channels from $N$ to $E$.

Then, the stacked 1--D (or temporal) dilated convolutional blocks~\cite{ref_tcn}~\cite{ref_convtasnet} are used for near-end speech estimation by concatenating the near-end and the mixture spectrograms as inputs. Each layer consists of 1--D convolutional blocks with increasing dilation factors as is denoted in Fig.~\ref{fig:structure}(a) with different colors. $M$ convolutional blocks with dilation factors $1,2,4, \ldots, 2^{M-1}$ are repeated $R$ times, obtaining a receptive field of $r(R,M,L) = R * 2^M * L$. The input to each block is zero padded accordingly to ensure the same output length as the input. For each 1--D convolutional block, a residual path and a skip-connection path are applied: the residual path of a block serves as the input to the next block, and the skip-connection paths of all blocks are concatenated for layer attention. Depthwise separable convolution is used, which is realized by a depthwise convolution ($D$-$Conv(\cdot)$) followed by a pointwise convolution ($1$$\times $$1$-$Conv(\cdot)$). The parametric rectified linear unit (PReLU) activation function~\cite{ref_prelu} and a normalization operation are performed after both first $1$$\times $$1$-$Conv(\cdot)$ and $D$-$Conv(\cdot)$ blocks respectively. The normalization can be chosen as global layer normalization (gLN) for noncausal configuration and cumulative layer normalization (cLN) for causal configuration~\cite{ref_convtasnet}\cite{ref_layernorm}.

Given the signal from the previous layer $\mathbf{e}^{(j-1)}  \in \mathbb{R}^{{T} \times 1 \times S}$, we compute activations $\mathbf{e}^{(j)}  \in \mathbb{R}^{{T} \times 1 \times S}$ with

\begin{equation}
\mathbf{e}^{(j)}=f\left(W^{(j)} * \mathbf{e}^{(j-1)}+b\right)
\end{equation}
where $W^{(j)}$ is the convolution kernel of the $j$-th layer, $f(\cdot)$ is the the activation function, $*$ is the convolution operator, $b$ is a bias. The vectors from all layers are expressed as $\mathbf{E} = \{\mathbf{e}^{(1)}, \mathbf{e}^{(2)}, ..., \mathbf{e}^{(j)}, ..., \mathbf{e}^{(J)}\} \in \mathbb{R}^{{T} \times J \times S}$, where $J = M * R$.

$\textbf{Layer Attention}$
Attention mechanism is performed here among multi-scale features from different layers to select important features. It is done as follows: The representation from the previous hidden state of the encoder LSTM unit, denoted by $\mathbf{Q} \in \mathbb{R}^{{T} \times 1 \times S}$ (performed as the $query$ of the attention) is dot-product with the multi-scale features from all layers (performed as $key$ and $value$ of the attention, and totally $J$ layers), generating attention weights. The $value$ are weighted averaged using the attention weights, and then fed to an encoder constructed by LSTM for near-end speech estimation. Multi-head attention~\cite{ref_attention} is used and formulated as,
\begin{equation}
\begin{array}{c}
\mathbf{Q}_{i}=\mathbf{Q} \mathbf{W}_{i}^{\mathbf{Q}}, \mathbf{K}_{i}=\mathbf{K} \mathbf{W}_{i}^{\mathbf{K}}, \mathbf{V}_{i}=\mathbf{V} \mathbf{W}_{i}^{\mathbf{V}}, \quad i=1, \cdots, h \\
\quad \text {head}_{i}=\text {Attention}\left(\mathbf{Q}_{i}, \mathbf{K}_{i}, \mathbf{V}_{i}\right), \quad i=1, \cdots, h \\
\text {Attention} (\mathbf{Q}_i, \mathbf{K}_i, \mathbf{V}_i)=\operatorname{softmax}\left(\frac{\mathbf{Q}_i \mathbf{K}_i^{T}}{\sqrt{d_{k}}}\right) \mathbf{V}_i \\
\operatorname{MultiHead}(\mathbf{Q}, \mathbf{K}, \mathbf{V})=\operatorname{Concat}\left(\operatorname{head}_{1}, \cdots, \text {head}_{h}\right) \mathbf{W}^{O}
\end{array}
\end{equation}
where $\mathbf{Q}$, $\mathbf{K}$, $\mathbf{V}$ are the query, the key and the value of the attention, and $\mathbf{K} = \mathbf{V} = \mathbf{E}$. $\mathbf{W}_{i}^{\mathbf{Q}} \in \mathbb{R}^{S \times F}$, $\mathbf{W}_{i}^{\mathbf{K}} \in \mathbb{R}^{S \times F}$, $\mathbf{W}_{i}^{\mathbf{V}} \in \mathbb{R}^{S \times F}$ are the projection matrix of the query, the key and the value for $i$-th head, $d_{k}=S$ is the dimension of the key, $\mathbf{W}^{O} \in \mathbb{R}^{F \times S}$ are the output matrix. To keep the number of attention parameters not increasing with the number of heads, compact attention is used$\footnote{Compact attention used in this paper was realied by: \url{ https://github.com/CyberZHG/torch-multi-head-attention}}$. An intuitive understanding of the layer attention mechanism is shown in Fig.~\ref{fig:structure}(c) by visualizing the attention weights of an audio clip example. It seems that the network has learned a special pattern. The network hyperparameters are described in Table~\ref{tab:configuration}.

\begin{table}[htb]
\setlength{\abovecaptionskip}{0.2cm}
\setlength{\belowcaptionskip}{0.2cm}
\centering
\caption{Hyperparameters of the network}
\label{tab:configuration}
\begin{tabular}{c|c}
\hline
\textbf{Symbol} & \textbf{Description} \\ \hline \hline
$T$ & Time steps of the spectrogram \\ \hline
$N$ & Number of filters in encoder and decoder \\ \hline
$L$ & Length of the filters (in samples) \\ \hline
\multirow{2}{*}{$E$} & Number of channels in bottleneck and \\
  & the residual paths' 1--D conv block \\ \hline
\multirow{2}{*}{$S$} & Number of channels in skip-connection \\
                    & paths' of 1--D conv block \\ \hline
$H$ & Number of channels in convolutional blocks \\ \hline
$K$ & Kernel size in convolutional blocks \\ \hline
$M$ & Number of convolutional blocks in each repeat \\ \hline
$R$ & Number of repeats \\ \hline
$F$ & Projection hidden size of the attention \\ \hline
$h$ & Number of the attention head \\ \hline
$\text{Causal}$ & Causal (true) or noncausal (false) \\ \hline
\end{tabular}
\end{table}

\subsection{Masking}
The estimated near-end representation is mapped to masks by $1$$\times $$1$-$Conv(\cdot)$ with Sigmoid activation functions. The $1$$\times $$1$-$Conv(\cdot)$ converts the number of channels back to that of the encoder.

\subsection{Loss Function}
\label{sec:loss_cun}
The loss function is calculated using the mean square error (MSE) of the estimated near-end waveform and the corresponding clean near-end waveform, formulated by,
\begin{equation} \label{eq:mask_loss}
{Loss_{\rm{MSE}}} = \sum_{n}\left(\hat{s}(n)-s(n)\right)^{2} \\
\end{equation}
where ${Loss_{\rm{MSE}}}$ is the MSE between the estimated near-end waveform $\hat{s}(n)$ and the raw one ${s}(n)$, $n$ is the time step.

\section{Experiments}
\label{sec:experiment}

\subsection{Performance metrics}
Two metrics are evaluated here, i.e., the echo return loss enhancement (ERLE) \cite{erle} for single-talk periods (periods without near-end signal) and the perceptual evaluation of speech quality (PESQ) for double-talk periods. ERLE reveals the echo attenuation achieved by the system. PESQ is obtained by comparing the estimated near-end speech with the clean one$\footnote{PESQ script: \url{https://www.itu.int/rec/T-REC-P.862/en}}$.

\subsection{Data preparation}
Since TIMIT dataset was widely used to evaluate AEC performance, we follows the data preparation method as same with \cite{DeLiang,Fazel,Fazel1}, resulting in 3500 training mixtures, and 300 test mixtures. Here, the RIR (Room Impulse Response) is used for simulating room reflections~\cite{rir}, and the NLP (Non-Linear Processing) is used for emulating the distortion introduced by loudspeaker. The far-end waveform is first distorted by the NLP, and then convoluted with the RIR, obtaining the echo waveform. It is then added to the near-end waveform to obtain the mixture. The mixture is then added by a noise waveform. The far-end and the mixture waveform were fed to the network as inputs, and the corresponding clean near-end waveform was considered as the estimated objective of the network.

The RIR and the NLP in the experiments are configured as is referred to in \cite{DeLiang,Fazel,Fazel1}, resulting in 7 RIRs, of which the first 6 RIRs are used to generate training mixtures and the last one is used to generate test mixtures. The hard clipping is used to simulate the power amplifier of loudspeaker, and the memoryless sigmoidal function is applied to emulate the nonlinear characteristic of loudspeaker, resulting in $x_{nl}(n)$ for nonlinear inputs.

Finally, the linear model $y_{lin}(n)$ and nonlinear model $y_{nl}(n)$ of acoustic path are obtained by convolving the linear input $x(n)$ and the non-linear input $x_{nl}(n)$ with a randomly chosen RIR $g(n)$,
\begin{equation}
\setlength{\abovedisplayskip}{3pt}
\setlength{\belowdisplayskip}{3pt}
\begin{array}{l}
y_{lin}(n) = x(n) * g(n),\quad y_{nl}(n) = x_{nl}(n) * g(n)
\end{array}
\end{equation}
where $*$ indicates the convolution manipulation.

For training, we generated the mixtures at signal to echo ratio (SER)~\cite{DeLiang} randomly chosen from {-6, -3, 0, 3, 6} dB. For testing, we generated the mixtures at three different SER {0, 3.5, and 7} dB.

\subsection{Model configurations}
In our experiments, waveforms at 16 kHz sample rate were directly served as the inputs. $\emph{Adm}$ algorithm was used for training with an exponential learning rate decaying strategy, where the learning rate starts at $1$$\times$$10^{-4}$ and ends at $1$$\times$$10^{-8}$. The total number of epochs was set to be 200. The criteria for early stopping is no decrease in the loss function on validation set for 10 epochs. The other parameters listed in Table~\ref{tab:configuration} are: $L$$=$$40$, $N$$=$$512$, $E$$=$$128$, $S$$=$$128$, $H$$=$$256$, $K$$=$$3$, $X$$=$$8$, $R$$=$$3$, $F$$=$$64$, $h$$=$$16$, $\text{Causal}$$=$$\text{True}$. The best number of attention heads was obtained at $h$$=$$16$ through plenty of trials. And, the reverberation time of $r(R,M,L) = L * R * 2^M = 1.92s$ can be handled. Other configurations can also be set according to the application scenarios.

\subsection{Results}

We first evaluated our method using linear and nonlinear models of acoustic path, respectively. Three schemes were used for comparisons, i.e., frequency domain normalized least mean square (denoted by AES+RES) \cite{fdnlms},  bidirectional long short-term memory method (denoted by BLSTM) \cite{DeLiang}, deep multitask (denoted by Multitask) \cite{Fazel}. Table~\ref{tab:linear} shows the average ERLE values and PESQ gains for these schemes. The $\Delta$PESQ is calculated as the difference of PESQ value of each method with respect to its unprocessed PESQ. The results show that our method obtain higher PESQ performance compared with Multitask method both for linear and nonlinear models. The superiority is obvious. Though, the ERLE performance of our method is slightly lower than the Multitask method, we think it is inessential because this value is sufficient low no matter for human hearing or speech recognition. Moreover, some simple postprocessing methods can be used to further increase this value, for example, setting the corresponding audio frame to zeros if the absolute value of the audio amplitude is lower than a threshold, such as -50dB. It also shows that the proposed method outperforms both AES+RES and BLSTM methods in all conditions. Some audio clips of the experiments can be found in the repository$\footnote{\url{https://github.com/ROAD2018/multi_scale_att_aec}}$.

\begin{table}[htb]
\setlength{\abovecaptionskip}{0.2cm}
\setlength{\belowcaptionskip}{0.2cm}
\centering
\caption{Performance of simulated linear model}
\label{tab:linear}
\begin{tabular}{|c|c|c|c|c|}
\hline
\multirow{2}{*}{\textbf{Metrics}}     & \multirow{2}{*}{\textbf{Methods}} & \multicolumn{3}{c|}{\textbf{Testing SER (dB)}} \\ \cline{3-5}
                      &       & \textbf{0}     & \textbf{3.5}    & \textbf{7}           \\ \hline \hline
\multicolumn{5}{|c|}{\textbf{Linear RIR Model}}  \\ \hline \hline
\multirow{4}{*}{\textbf{ERLE (dB)}} & AES+RES                 & 29.38       & 25.88       & 21.97       \\ \cline{2-5}
                      & BLSTM    & 51.61   & 50.04  & 47.42       \\ \cline{2-5}
                      & Multitask   & 64.66    & 64.16   & 62.26    \\ \cline{2-5}
                      & Proposed   & 58.53   & 56.27  & 54.96      \\ \hline
\multirow{4}{*}{\textbf{$\Delta$ PESQ}} & AES+RES & 0.93 & 0.81 & 0.68 \\ \cline{2-5}
                      & BLSTM      & 0.8    & 0.78   & 0.74        \\ \cline{2-5}
                      & Multitask   & 1.04  & 1.02   & 0.99      \\ \cline{2-5}
                      & Proposed    & 1.76  & 1.74  & 1.56     \\ \hline
\hline
\multicolumn{5}{|c|}{\textbf{Nonlinear RIR Model}}  \\ \hline \hline
\multirow{3}{*}{\textbf{ERLE (dB)}} & AES+RES & 16.76 & 14.26 & 12.33 \\ \cline{2-5}
                      & Multitask     & 61.79   & 60.52   & 59.47  \\ \cline{2-5}
                      & Proposed     & 56.69   & 55.72  & 52.65  \\ \hline
\multirow{3}{*}{\textbf{$\Delta$ PESQ}} & AES+RES & 0.54 & 0.43 & 0.31 \\ \cline{2-5}
                      & Multitask       & 0.84   & 0.83    & 0.81  \\ \cline{2-5}
                      & Proposed       & 1.67    & 1.64   & 1.43     \\ \hline
\end{tabular}
\end{table}

Further, we also evaluated the performance in presence of additive noise and nonlinear model of acoustic path. When generating the training data, we added a white noise at 10dB SNR level with nonlinear acoustic path at 3.5dB SER level. We compared our method against AES+RES and the Multitask methods. As is shown in Table~\ref{tab:noisy}, our framework gains the best performance both for PESQ and ERLE. This means that our model has superior noise robustness which is more realistic in real-world scenarios.

\begin{table}[htb]
\setlength{\abovecaptionskip}{0.2cm}
\setlength{\belowcaptionskip}{0.2cm}
\centering
\caption{Performance for noisy scenario}
\label{tab:noisy}
\begin{tabular}{|c|c|c|c|c|}
\hline
\multirow{2}{*}{\textbf{Methods}}  & \multicolumn{2}{c|}{\textbf{Metrics}} \\ \cline{2-3}
& \textbf{ERLE (dB)}   & \textbf{\textbf{$\Delta$ PESQ}}  \\ \hline \hline
AES+RES    & 10.13   & 0.21    \\ \hline
Multitask   & 46.12 & 0.70     \\ \hline
Proposed   & 52.99  & 1.53  \\ \hline
\end{tabular}
\end{table}

Moreover, to evaluate our framework in realistic scenarios, real measured RIRs selected from the Aachen impulse response database \cite{realRIR} were used in the experiments. In this experiment, we used the real measured RIRs captured in ``office'', ``meeting room'', ``lecture room'', ``stairway1'', ``stairway2'', ``bathroom'', and ``lecture room'' for training and ``corridor'' in HHP for testing. Two schemes were compared as is listed in Table~\ref{tab:reallinear}, i.e., the AES+RES method and the context-aware method (denoted by CAD-AEC) \cite{Fazel1}. Our method gains higher PESQ compared with other methods. The ERLE gain of our model is a little lower than the CAD-AEC method, but higher than the AES+RES method.

\begin{table}[htb]
\setlength{\abovecaptionskip}{0.2cm}
\setlength{\belowcaptionskip}{0.2cm}
\centering
\caption{Performance of real measured linear model}
\label{tab:reallinear}
\begin{tabular}{|c|c|c|c|c|}
\hline
\multirow{2}{*}{\textbf{Metrics}}   & \multirow{2}{*}{\textbf{Methods}} & \multicolumn{3}{c|}{\textbf{Testing SER (dB)}} \\ \cline{3-5}
                      &       & \textbf{0}     & \textbf{3.5}    & \textbf{7}           \\ \hline \hline
\multirow{3}{*}{\textbf{ERLE (dB)}} & AES+RES    & 12.16    & 11.46    & 10.52       \\ \cline{2-5}
                      & CAD-AEC     & 56.51   & 60.49   & 61.39  \\ \cline{2-5}
                      & Proposed    & 55.74   & 54.41   & 51.84  \\ \hline
\multirow{3}{*}{\textbf{$\Delta$ PESQ}} & AES+RES   & 0.57   & 0.53    & 0.48       \\ \cline{2-5}
                      & CAD-AEC       & 1.11   & 1.06    & 1.00  \\ \cline{2-5}
                      & Proposed      & 1.46  & 1.48   &  1.33     \\ \hline
\end{tabular}
\end{table}
Finally, to evaluate the performance when the training and testing conditions are more different than the previous experiments, we generated seven synthetic RIRs for training and again tested on data that was created by the real measured ``corridor'' RIRs which is deployed as depicted in the Fig. 9(c) in \cite{realRIR}. The reverberation time ($T_{60}$) is matched with the `corridor'' environment which is selected from $\{0.2, 0.4, 0.6, 0.8, 0.9, 1.0, 1.25\}$s. We again compared the results of our method against the CAD-AEC method \cite{Fazel1} both in linear and nonlinear RIR models.
As is revealed in~Table \ref{tab:generalization}, for both linear and nonlinear RIR models, our method outperforms the CAD-AEC method in terms of the PESQ gain. Our model gains higher ERLE in nonlinear model but lower in linear model. This reveals the superior nonlinear modelling ability of our model.

\begin{table}[htb]
\setlength{\abovecaptionskip}{0.2cm}
\setlength{\belowcaptionskip}{0.2cm}
\centering
\caption{Training on synthetic RIRs and testing on real RIRs.}
\label{tab:generalization}
\begin{tabular}{|c|c|c|c|c|}
\hline
\multirow{2}{*}{\textbf{Metrics}}   & \multirow{2}{*}{\textbf{Methods}} & \multicolumn{3}{l|}{\textbf{Testing SER (dB)}} \\ \cline{3-5}
                      &       & \textbf{0}     & \textbf{3.5}    & \textbf{7}           \\ \hline \hline
\multicolumn{5}{|c|}{\textbf{Linear RIR Model}}  \\ \hline \hline
\multirow{2}{*}{\textbf{ERLE}} & CAD-AEC & 42.66   & 47.96   & 52.47  \\ \cline{2-5}
                      & Proposed  & 45.78    & 43.36    & 42.22       \\ \hline
\multirow{2}{*}{\textbf{$\Delta$ PESQ}} & CAD-AEC   & 0.90  & 0.82  & 0.73  \\ \cline{2-5}
                      & Proposed  & 1.18   & 1.19   & 1.08  \\ \hline \hline
\multicolumn{5}{|c|}{\textbf{Nonlinear RIR Model}}  \\ \hline \hline
\multirow{2}{*}{\textbf{ERLE}} & CAD-AEC & 19.08   & 19.97   & 21.64  \\ \cline{2-5}
                      & Proposed  & 38.63   & 36.66  & 34.71  \\ \hline
\multirow{2}{*}{\textbf{$\Delta$ PESQ}} & CAD-AEC  & 0.95   & 0.90    & 0.82  \\ \cline{2-5}
                      & Proposed  & 1.27   & 1.23   & 1.04   \\ \hline
\end{tabular}
\end{table}

\section{Conclusions and discussions}
We proposed an end-to-end neural network with layer attention for acoustic echo cancellation. Temporal convolution network with stacked dilated convolutions was used to extract multi-scale representations from the spectrograms of the far-end and the mixture. Attention was used among these representations to highlight important features.
The superiority of the proposed framework in double-talk, background noise, and nonlinear distortion scenarios were validated through experiments using simulated and real-recorded RIRs. The generality of the model was also tested by training on simulated RIRs and validating on real-recorded RIRs.

The receptive field of the network can be configured to cover the reverberation time and the system delay. Therefore, a data buffer is required to cover this receptive field. The buffer is initialized with zeros at the beginning of signal processing, and then a new frame is fed into the buffer and the old one slides out the buffer. In this way, our method can be implemented in real-time scenarios.

\bibliographystyle{IEEEtran}
\bibliography{refs21}

\end{sloppy}
\end{document}